\documentclass[twocolumn]{aastex6}
\pdfoutput=1
\usepackage{graphics,graphicx,rotating,amsmath}
\usepackage{natbib}
\defcitealias{SCpaper}{W14}
\defcitealias{Zwaan2010}{Z10}
\defcitealias{Bekeraite2016}{B16}
\defcitealias{Papastergis2011}{P11}
\bibpunct{(}{)}{;}{a}{}{,} 
\usepackage{flexisym}
\usepackage{textcomp}
\usepackage{enumerate}
\usepackage{listings}
\usepackage{url}
\usepackage{lineno}
\shortauthors{Bekerait\.{e} et al. }
\shorttitle{The CALIFA and HIPASS circular velocity function}
\fullcollaborationName{The CALIFA Collaboration}
\received{May 11th}
\revised{July 22th}
\accepted{August 1st}
\begin{document} 

\title{THE CALIFA AND HIPASS CIRCULAR VELOCITY FUNCTION FOR ALL MORPHOLOGICAL GALAXY TYPES}
\author{S.~Bekerait{\.e}\altaffilmark{1}, C.J.~Walcher\altaffilmark{1}, L.~Wisotzki\altaffilmark{1}, D.J.~Croton\altaffilmark{2}, J.~Falc{\'o}n-Barroso\altaffilmark{3, 4}, M.~Lyubenova\altaffilmark{5}, D.~Obreschkow\altaffilmark{6}, S.~F.~S\'anchez\altaffilmark{7}, K.~Spekkens\altaffilmark{8}, P.~Torrey\altaffilmark{9, 10}, G.~van de Ven\altaffilmark{11}, M.A. Zwaan\altaffilmark{12}, Y.~Ascasibar\altaffilmark{13}, J.~Bland-Hawthorn\altaffilmark{14}, R.~Gonz\'alez Delgado\altaffilmark{15}, B.~Husemann\altaffilmark{16}, R.A.~Marino\altaffilmark{17, 18}, M.~Vogelsberger\altaffilmark{19}, B. Ziegler\altaffilmark{20}}

\altaffiltext{1}{Leibniz-Institut f\"ur Astrophysik Potsdam (AIP), An der Sternwarte 16, D-14482 Potsdam, Germany}
\altaffiltext{2}{Centre for Astrophysics and Supercomputing, Swinburne University of Technology, Hawthorn, Victoria 3122, Australia}
\altaffiltext{3}{Dept. Astrof\'{i}sica, Universidad de La Laguna, C/ Astrof\'{i}sico Francisco S\'anchez, E-38205-La Laguna, Tenerife, Spain\label{i:IAC}}
\altaffiltext{4}{Instituto de Astrof\'{i}sica de Canarias, C/V\'{\i}a L\'actea S/N, 38200-La Laguna, Tenerife, Spain\label{i:ULL}}
\altaffiltext{5}{Kapteyn Astronomical Institute, University of Groningen, Postbus 800, 9700 AV Groningen, The Netherlands\label{i:Kapteyn}}
\altaffiltext{6}{ICRAR, University of Western Australia, 35 Stirling Highway, Crawley, WA 6009, Australia\label{i:ICRAR}}
\altaffiltext{7}{Instituto de Astronom\'ia, Universidad Nacional Auton\'oma de M\'exico, A.P. 70-264, 04510, M\'exico, D.F.\label{i:UNAM}}
\altaffiltext{8}{Department of Physics, Royal Military College of Canada, P.O. Box 17000, Station Forces, Kingston, ON, K7K 7B4, Canada\label{i:RMCC}}
\altaffiltext{9}{Department of Physics, Kavli Institute for Astrophysics and Space Research, Massachusetts Institute of Technology, Cambridge, MA 02139, USA}
\altaffiltext{10}{TAPIR, Mailcode 350-17, California Institute of Technology, Pasadena, CA 91125, USA}
\altaffiltext{11}{Max Planck Institute for Astronomy, K\"onigstuhl 17, D-69117 Heidelberg, Germany\label{i:MPIA}}
\altaffiltext{12}{ESO, ALMA Regional Centre, Karl-Schwarzschild-Strasse 2, D-85748 Garching, Germany\label{i:ESO_ALMA}}
\altaffiltext{13}{Departamento de F\'isica Te\'orica, Facultad de Ciencias, Universidad Aut\'onoma de Madrid, E-28049 Madrid, Spain\label{i:UAM}}
\altaffiltext{14}{Sydney Institute for Astronomy, School of Physics, University of Sydney, NSW 2006, Australia\label{i:Sydney}}
\altaffiltext{15}{Instituto de Astrof\'isica de Andaluc\'ia (IAA/CSIC), Glorieta de la Astronom\'{\i}a s/n Aptdo. 3004, E-18080 Granada, Spain\label{i:IAA}}
\altaffiltext{16}{European Southern Observatory, Karl-Schwarzschild-Str. 2, D-85748 Garching b. M{\"u}nchen, Germany\label{i:ESO}}
\altaffiltext{17}{Department of Physics, Institute for Astronomy, ETH Z{\"u}rich, CH-8093 Z{\"u}rich, Switzerland\label{i:ETH}}
\altaffiltext{18}{Departamento de Astrof\'{i}sica y CC. de la Atm\'{o}sfera, Facultad de CC. F\'{i}sicas, Universidad Complutense de Madrid, Avda. Complutense s/n, 28040 Madrid, Spain\label{i:Madrid}}
\altaffiltext{19}{Department of Physics, Kavli Institute for Astrophysics and Space Research, Massachusetts Institute of Technology, Cambridge, MA 02139, USA}
\altaffiltext{20}{University of Vienna, T\"urkenschanzstr. 17, 1180 Vienna, Austria \label{i:vienna}
}

\begin{abstract}
The velocity function is a fundamental observable statistic of the galaxy population, similarly important as the luminosity function, but much more difficult to measure. In this work we present the first directly measured circular velocity function that is representative between 60 $< v_{\mathrm{circ}} <$ 320 km s$^{-1}$ for galaxies of all morphological types at a given rotation velocity. For the low mass galaxy population (60 $< v_{\mathrm{circ}} <$ 170  km s$^{-1}$), we use the HIPASS velocity function. For the massive galaxy population (170 $< v_{\mathrm{circ}} <$ 320 km s$^{-1}$), we use stellar circular velocities from the Calar Alto Legacy Integral Field Area Survey (CALIFA). In earlier work we obtained the measurements of circular velocity at the 80\% light radius for 226 galaxies and demonstrated that the CALIFA sample can produce volume-corrected galaxy distribution functions. The CALIFA velocity function includes homogeneous velocity measurements of both late and early-type rotation-supported galaxies and has the crucial advantage of not missing gas-poor massive ellipticals that HI surveys are blind to. We show that both velocity functions can be combined in a seamless manner, as their ranges of validity overlap. The resulting observed velocity function is compared to velocity functions derived from cosmological simulations of the $z =$ 0 galaxy population. We find that dark matter-only simulations show a strong mismatch with the observed VF. Hydrodynamic simulations fare better, but still do not fully reproduce observations.
\end{abstract}

\section{Introduction}
The circular velocity function (VF), the space density of galaxies as a function of their \added{circular} rotation velocities, is directly related to total dynamical masses of the galaxies and not dominated by their baryonic content, unlike the galaxy luminosity function (LF) \citep{Desai2004}. As a tracer of dark matter halo masses \citep[hereafter Z10]{Zwaan2010}, the VF can be used as a test of the $\Lambda$CDM paradigm \citep{Klypin2014, Papastergis2011} and a probe of cosmological parameters \citep{Newman2000, Newman2002} or the relation between the dark matter halo and galaxy rotation velocities.

Observationally, VF differs significantly from the LF. The latter, although difficult to predict and interpret theoretically, is much easier to measure \citep{Klypin2014} \added{and does not depend on the spatial distribution of baryons in galaxies. Depending on the precise definition of circular velocity, the VF is a function of both the halo and baryonic mass spatial distribution and their ratio in a particular galaxy, however, it is not significantly affected by uncertainties in the stellar mass-to-light ratio. In this regard it is a superior tool for testing the results of cosmological simulations}. 

Measuring the VF is difficult on all halo mass scales. Cluster rotation velocities have completely different dynamical properties and require different observational methods than individual galaxies \citep{Kochanek2001}, while the lowest velocity galaxy samples are not complete. Even at intermediate\replaced{-high masses}{velocities} the VF has not been fully constrained, because circular velocity measurements for gas-poor early-types, dominating the high velocity end of the galaxy velocity function, are notoriously challenging \citep{Gonzalez2000, Papastergis2011, Obreschkow2013CDM}. \added{Moreover, even though the circular velocity is easy to define theoretically, there is no clear observational definition, especially given that the rotation curves of some classes of galaxies do not flatten}.

Several studies have attempted to use galaxy scaling relations in order to infer circular velocities from more accessible observable quantities. \citet{Gonzalez2000} estimate a VF by converting the SSRS2 luminosity function using the Tully-Fisher relation. \deleted{They include ellipticals by converting their velocity dispersions to circular velocity estimates}. A similar approach was adopted by \citet{Abramson2014}, who construct galaxy group and field VFs using velocity estimates based on Sloan Digital Sky Survey (SDSS) photometric data. \citet{Desai2004} construct cluster and field VFs by using SDSS \deleted{Early Data Release} data\replaced{ and}{,} Tully-Fisher and Fundamental Plane relations\deleted{, also accounting for the baryon impact on the rotation curves and the scatter in the scaling relations used in their work}.

In \citet{Klypin2014} a Local Volume VF, complete down to $v_{\mathrm{circ}}$ $\approx$ 15 km s$^{-1}$, was estimated using a combination of HI observations and line-of-sight velocities estimated from photometry. However, their study does not sample the velocities above $\approx$ 200 km s$^{-1}$. 

An HI velocity function down to 30 km s$^{-1}$ was directly measured from HI Parkes All Sky Survey (HIPASS) linewidths \citepalias{Zwaan2010}. Nevertheless, as shown in \cite{Obreschkow2013CDM}, massive, rapidly rotating, gas-poor ellipticals are systematically missing from HIPASS data. Therefore its high velocity end is very incomplete. 

\citet{Papastergis2011} (P11) estimate the HI linewidth function from Arecibo Legacy Fast ALFA (ALFALFA) survey data and suggest using the linewidth function as a more useful probe of the halo mass distribusion. They also provide a VF for all types by combining their VF with the velocity function converted from \citet{Chae2010} velocity dispersion measurements. \deleted{Finally, they derive a late-type VF from ALFALFA linewidth function assuming random orientation of galaxies.}

CALIFA is in a unique position with its well-understood selection function \replaced{ and}{,} a wide field of view \replaced{. In addition, the CALIFA sample is}{and} the first IFS sample that includes a large number of galaxies with diverse morphologies. As described in \citet{Krajnovic2008}, 80\% of early-type galaxies can be expected to have a rotating component\deleted{ and contribute to the VF}. The use of stellar kinematics enables us to include gas-poor, rotating early-type galaxies in a homogeneous manner. Therefore we are able to directly measure the VF for rotating galaxies of all morphological types, in contrast to the inferred VFs reported by \citet{Gonzalez2000, Desai2004, Chae2010, Abramson2014}.

Within this work, we assume a benchmark cosmological model with $H_0 =$ 70 km s$^{-1}$/Mpc, $\Omega_{\Lambda} =$ 0.7 and $\Omega_{\mathrm{M}} =$ 0.3. All VFs from the literature were rescaled to this particular cosmology, as described in \citet{Croton2013}.

\section{CALIFA stellar circular velocity measurements}
\label{sec:vel_measurements}

CALIFA observations use the PMAS instrument \citep{Roth2005} in PPaK \citep{Verheijen2004} mode, mounted on the 3.5 m telescope at the Calar Alto observatory. The CALIFA survey, sample and data analysis pipeline are described in detail \citet{Sanchez2012, Husemann2013, Garcia-Benito2015, SCpaper}. We refer the reader to the first paper of the CALIFA stellar kinematics series (Falc{\'o}n-Barroso et al., submitted) where the kinematic map extraction and sample is described in full detail. 

In this analysis, we use the "useful" galaxy sample defined in \citet[B16]{Bekeraite2016} and the circular velocity measurements obtained therein. Briefly, we start with the initial \added{statistically complete} sample of 277 available stellar velocity fields, 51 of which were not useful for further analysis due to S/N issues (low number of Voronoi bins, foreground contamination) or extremely distorted velocity fields. The final sample consisted of 226 galaxies. As shown in \citetalias{Bekeraite2016}, the rejected galaxies were predominantly fainter (SDSS $M_r > -20$ mag), which affected the lower completeness limit but did not introduce bias in the sample above it. 

We then fit the position and rotation curve parameters by performing Markov Chain Monte Carlo (MCMC) modelling of the \deleted{2D} velocity fields. The rotation velocity $v_{\mathrm{opt}}$ was estimated by evaluating the model rotation velocity at the 80\% light radius (the optical radius). Due to CALIFA's large but still limited field of view rotation curve extrapolation was necessary for 165 galaxies. \deleted{The majority of the sample (134 galaxies) were sampled at least to 80\% of $r_{\mathrm{opt}}$, as shown in \citetalias{Bekeraite2016}. Only a single galaxy was not sampled up to one effective radius.}

We do not split the galaxy sample into ellipticals and spirals to estimate their $v_{\mathrm{circ}}$ values separately. Instead, as described in Sec. 4.4 of \citetalias{Bekeraite2016}, a correction estimated in \citet{Kalinova2016} has been applied to all galaxies. \citet{Kalinova2016} analyse the relationship between dynamical masses inferred using \replaced{classical ADC models}{the classical ADC approach} \added{\citep[see Chapter 4,][]{BinneyTremaine}} and axisymmetric Jeans anisotropic Multi-Gaussian (JAM) models applied to stellar mean velocity and velocity dispersion fields of 18 late-type galaxies observed
with the SAURON IFS instrument. We utilise the relation provided in their Table 4 and calculate the circular velocities by multiplying the measured velocity by the square root of the factors provided, based on the ratio between the $v_{\mathrm{opt}}$ and the line-of-sight stellar velocity dispersion at the optical radius. We demonstrate that the obtained circular velocity is comparable with ionised gas rotation velocity in \citetalias{Bekeraite2016}.

\section{Results}
\subsection{CALIFA circular velocity function}
\label{sec:CALIFA_VF}
We measure the CALIFA stellar circular VF $\Psi_{\mathrm{circ}}$ in the same manner as the LFs in \citet[W14]{SCpaper} and \citetalias{Bekeraite2016}, estimating the optimal number of velocity bins using Scott's Rule \citep{Scott1979}. The $1/V_{\mathrm{max}}$ weights, corrected for cosmic variance as described in \citetalias{SCpaper} are assigned to each galaxy and then used to calculate the VF. We note that the uncertainties correspond to Poissonian errors in each bin only and do not include any uncertainties in circular velocity measurements \replaced{. The impact of measurement uncertainties is evaluated in Sec. \ref{sec:uncertainties}}{see Sec. \ref{sec:uncertainties}}.

\begin{figure*}[htb!]
\includegraphics{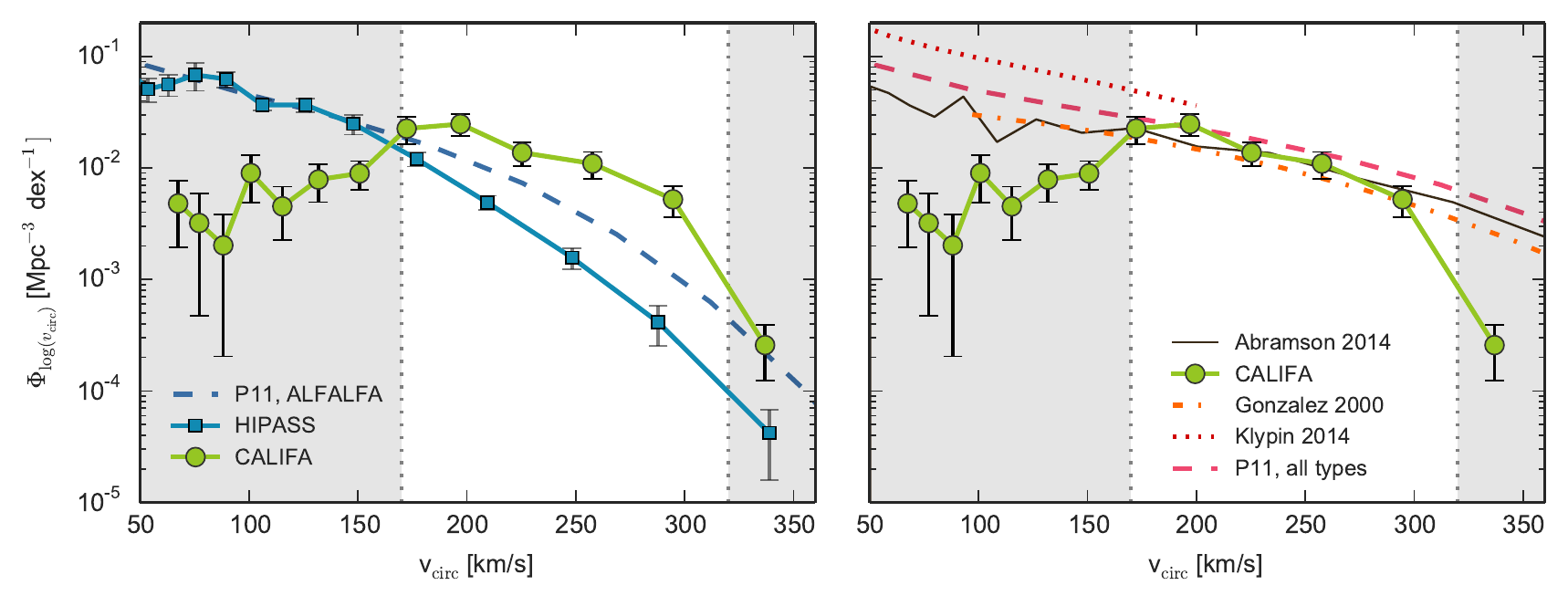} 
\caption{CALIFA velocity function, compared with the HIPASS \citepalias{Zwaan2010}, \citet{Gonzalez2000, Klypin2014, Abramson2014} and \citetalias{Papastergis2011} measurements. The left panel shows the comparison with the observed VFs of rotation-dominated gas-rich galaxies. The right panel displays the comparison with indirectly estimated VFs. The shaded areas and dotted vertical lines show an approximate region where incompleteness in the CALIFA sample becomes important\deleted{, estimated from the LF via the SDSS $r$-band Tully-Fisher relation}.}
\label{fig:vcirc_func}
\end{figure*}

As can be seen in Fig.\ref{fig:vcirc_func}, the high-velocity end of $\Psi_{\mathrm{circ}}$ lies significantly above the \citetalias{Zwaan2010} or \citetalias{Papastergis2011} HI velocity functions. This was to be expected since HI surveys are blind to gas-poor massive ellipticals \citep{Obreschkow2013CDM}. However, the CALIFA VF is below the higher-velocity end of \citetalias{Papastergis2011} inferred VF for all galaxy types. This is not surprising given that their VF combines the observed ALFALFA VF and the velocity dispersion function of \citet{Chae2010}. \replaced{In contrast,}{Our circular} VF is defined for rotation-dominated galaxies only and we do not include the velocity dispersion contribution in any way, barring the circular velocity correction described in Sec. \ref{sec:vel_measurements}. 
  
At lower velocities the CALIFA VF starts to fall off rapidly and deviates from the Schechter function shape. We estimate the region where incompleteness should become important, based on the luminosity function of the sample provided in \citetalias{Bekeraite2016}. We convert the luminosity completeness limits to velocity using the Tully-Fisher relation measured in \citetalias{Bekeraite2016} and find that the CALIFA VF should be complete within the velocity range of 140 $< v_{\mathrm{circ}} < $ 345 km s$^{-1}$. Such a direct conversion excludes the scatter in TFR, which causes a fall-off sooner than would be naively expected from the TFR alone. By taking the $rms$ TFR scatter (0.27 mag) into account, we find that the CALIFA velocity function, as shown in Fig.\ref{fig:vcirc_func}, can be safely assumed to be complete above 170 km s$^{-1}$. At the high velocity end, the CALIFA survey is limited by its survey volume as the total number of galaxies brighter than $M_r =$ -23 expected within the survey is of the order of unity. Given the low number of galaxies at the high velocity end of the TFR, we are unable to estimate the real TFR scatter among the most massive galaxies and the subsequent onset of bias. However, conservatively adopting the $rms$ TFR scatter of 0.27 mag we find that the CALIFA VF is complete at least up to 320 km s$^{-1}$.

\subsection{Uncertainties}
\label{sec:uncertainties}
The CALIFA stellar rotation velocity measurements have significant uncertainties, resulting from limited spatial resolution of binned stellar velocity fields, limited CALIFA field of view and pressure-support dependent correction term. The circular VF is likely affected by all these factors. A broader discussion of uncertainties in the circular velocity measurements and volume correction weights is contained in \citetalias{Bekeraite2016} and \citetalias{SCpaper}.

\begin{figure}[htbp!]
\includegraphics[width=\linewidth]{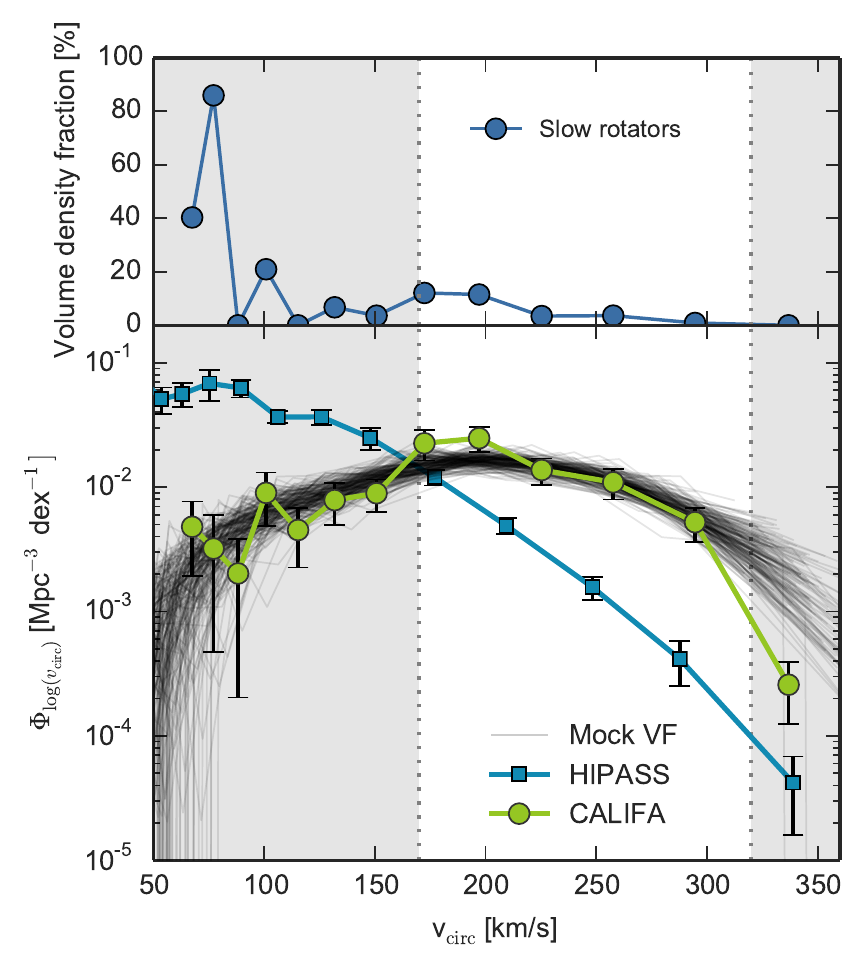} 
\caption{Top panel shows the volume density fraction of slow rotators (SR), for which the measured rotation velocities and circular velocity corrections are the most uncertain. This fraction does not reach \replaced{15\% of the volume densities at any velocity}{20\% of volume density at for $v_{\mathrm{circ}}$  $\gtrsim$ 110 km s$^-1$}.
Bottom panel shows the effect of velocity measurement uncertainties on the velocity function. Thin grey lines are the mock realisations of the VF. The green points and line are the CALIFA VF. The blue line shows the HIPASS VF.}
\label{fig:vcirc_func_test}
\end{figure}

In order to check the impact of velocity measurement uncertainties we employ a resampling method similar to \citetalias{Papastergis2011}. We generate 200 mock CALIFA VF samples (shown in Fig. \ref{fig:vcirc_func_test}) in which the volume weights are not changed, but the velocities $v_{\mathrm{circ}}$ are replaced with randomly drawn values such that $v^{\mathrm{test}}_{\mathrm{circ}}$ = $v_{\mathrm{circ}}$ + $\mathcal{N}(0, \sigma_v)$, where $\sigma_v$ are the individual velocity uncertainties of each point. 

Overall, the effect is a smoothing of the VF as the datapoints are 'smeared' into the neighbouring bins. As the 1/$V_{\mathrm{max}}$ weights are higher at the lower velocities, this leads to an artificial boost at the highest velocity end. Undoubtedly, this effect should be present in our VF as well, making the location of the highest velocity CALIFA datapoint even more uncertain. Given that this bin only includes 3 galaxies and is outside our estimated completeness range, we exclude it from all further analysis.

\subsection{Combined CALIFA-HIPASS circular velocity function}
In order to extend the VF to a wider velocity range we merge the HIPASS VF between 60-160 km s$^{-1}$ and CALIFA circular velocities between 160-320 km s$^{-1}$, effectively choosing the more complete VF in each bin. Merging the two VFs in this way is justified as HI-rich late-type galaxies dominate the counts below 200 km s$^{-1}$ \deleted{, where they are observed by HIPASS}. \replaced{In}{At} the high mass limit, early-type massive rotators contribute significantly to the high velocity end, where the CALIFA sample is expected to be complete at least up to $v_{\mathrm{circ}}=$ 320 km s$^{-1}$, as described above. 

\begin{figure}[htbp!]
\includegraphics[width=\linewidth]{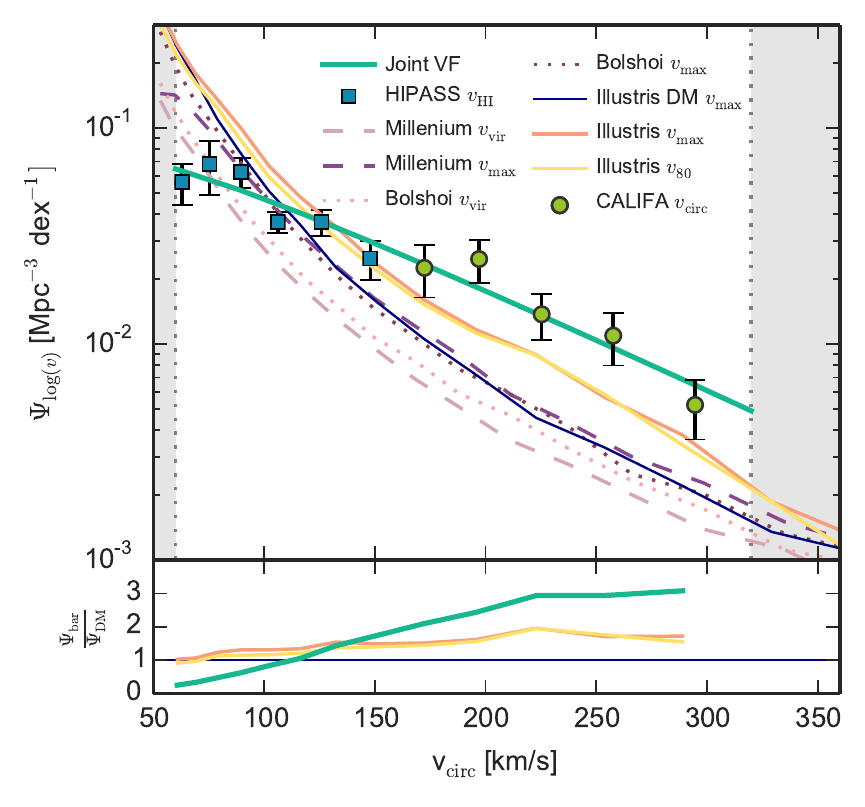} 
\caption{Zoomed-in view showing the combined CALIFA+HIPASS VF and the best Schechter fit, shown as thick solid teal line. 
In the top panel, dark matter-only halo VFs from Millenium and Bolshoi simulations are shown as purple and pink dashed and dotted lines. $v_{\mathrm{max}}$ VF from Illustris-1 dark-matter only simulation is shown by a thin dark blue line. Full-physics Illustris-1 simulation VFs, calculated using the subhalo $v_{\mathrm{max}}$ and \deleted{potential-induced} circular velocity at 80\% stellar mass radius ($v_{\mathrm{80}}$) are displayed as orange and yellow solid lines respectively. The lower panel shows the ratio between the baryonic simulated VFs, the combined CALIFA+HIPASS fit and the DM-only Illustris VF. See Sec. \ref{sec:sims} for a description, discussion, and references.}
\label{fig:combined_vcirc_func}
\end{figure}

We fit a Schechter function

\begin{equation} 
\label{eq:Schechter}
\Psi(v_{\mathrm{circ}}) = \Psi_{*}\left(\frac{v_{\mathrm{circ}}}{v_{*}}\right)^{\alpha}\exp\left[-\left(\frac{v_{\mathrm{circ}}}{v_{*}}\right)\right]
\end{equation}
 
\noindent to the combined VF, \deleted{which is }shown in Fig. \ref{fig:combined_vcirc_func}. The datapoints are listed in Table \ref{table:VF_values} and the fit parameters are provided in Table \ref{table:fit_VF}. Uncertainties for the HIPASS values were taken from \citet{Obreschkow2013CDM}, where they supplement direct measurement and shot noise with other uncertainties such as distance errors, cosmology uncertainties and cosmic variance. Similarly to \citetalias{Zwaan2010}, we find that the model parameters
are highly covariant. \\

\begin{table}
\caption{Schechter function fit parameters for CALIFA+HIPASS VF (Eq.\ref{eq:Schechter}).}             
\label{table:1}      
\centering                                      
\begin{tabular}{ c c c c}          
\hline\hline                        
 $\Psi_{*}$ [$\times10^{-3}$ Mpc$^{-3}$] & $v_{*}$ [km s$^{-1}$] & $\alpha$  \\    
\hline                                   
 130.0$\pm$35.8 & 89.3$\pm$32.8 & 0.2$\pm$0.6\\      
\hline                                             
\end{tabular}
\label{table:fit_VF}
\end{table}

\begin{table}
\caption{CALIFA-HIPASS velocity function values.}             
\label{table:2}      
\centering                                      
\begin{tabular}{c c c }          
\hline\hline                        
Survey & $v_{\mathrm{circ}}$ [km s$^{-1}$] & $\Psi(\log_{10}v_{\mathrm{circ}})$ [$\times10^{-3}$ Mpc$^{-3}$] \\    
\hline                                
& 63.0 & 56.1 $\pm$ 12.2 \\
& 75.3 & 68.2 $\pm$ 19.1 \\
HIPASS & 89.5 & 62.7 $\pm$ 10.1 \\
& 106.2 & 36.6 $\pm$ 4.1 \\
& 125.9 & 36.7 $\pm$ 5.1 \\
& 147.9 & 24.9 $\pm$ 5.0 \\

\hline 
& 172.4 & 22.5 $\pm$ 6.1 \\
& 197.1 & 24.7 $\pm$ 5.5 \\
CALIFA & 225.4 & 13.7 $\pm$ 3.3 \\
& 257.7 & 10.9 $\pm$ 3.0 \\
& 294.6 & 5.2 $\pm$ 1.6 \\
\hline                                             
\end{tabular}
\label{table:VF_values}
\end{table}

\subsection{Discussion}
Despite the care with which we have undertaken our analysis, combining the HI rotation velocities and stellar circular rotation velocities as we have done has some caveats. 

First of all, the actual methods used to construct the HIPASS and CALIFA VFs are different. The CALIFA volume correction procedure uses a more straightforward 1/V$_{\mathrm{max}}$ method \citep{Schmidt1968}\replaced{. In addition, as discussed in \citetalias{SCpaper}, the CALIFA volume weights are adjusted to account }{improved by accounting} for the radial density variations.

Meanwhile, \citetalias{Zwaan2010} employ a bivariate step-wise maximum likelihood (2DSWML) technique to obtain their space densities.
\citet{Zwaan2003} verify that the method is insensitive to even large radial density variations\deleted{, although they do find that small survey volume does affect the resulting distribution at the low mass end.} \replaced{However}{In addition}, the HIPASS linewidth function (WF) matches the WF obtained from the deeper ALFALFA survey down to 60 km s$^{-1}$ \citepalias{Papastergis2011}, confirming that the effect of large scale structure on the HIPASS VF is negligible, at least in the range of our analysis. As shown in \citet{Zwaan2003}, the 1/V$_{\mathrm{max}}$ and 2DSWML methods yield practically indistinguishable results, confirming that the two VFs derived using both methods are compatible. 

The HIPASS sample consists of late-type galaxies only, since visually classified early-type galaxies, comprising 11\% of the sample, have been removed. However, the fraction of early-type and S0 galaxies is reported to only have a noticeable effect on the VF only for galaxies with rotation velocities above 200 km s$^{-1}$, where we use CALIFA VF values already. \deleted{The non-rotational gas motions, also relatively more significant at lower velocity, were found not to have a significant effect on the HIPASS VF.} 

HIPASS linewidths have been corrected for inclination using SuperCOSMOS imaging $b$-band photometric axis ratios \citep{Meyer2008}, while we use kinematic inclinations obtained from MCMC modelling of the 2D velocity fields. Photometric inclination estimates are systematically affected by unknown intrinsic disk thickness, choice of $b/a$ measurement radius and any departure from a perfect circular disk shape. However, given that \citetalias{Zwaan2010} exclude galaxies with estimated inclinations $i <$ 45\textdegree \added{due to larger uncertainties at low inclinations}, inconsistencies in inclination measurements are unlikely to have had a significant effect.

\deleted{The HIPASS rotational velocities were derived from HI linewidths measured at 50 per cent peak flux and corrected for inclination, instrumental and relativistic broadening and turbulent motions, as described in \citet{Meyer2008}. Therefore, the CALIFA and HIPASS circular rotation velocities should be compatible as both have been corrected for noncircular motions and therefore are expected to trace the dynamical masses of the galaxies. }

As discussed in \citetalias{Zwaan2010}, HIPASS \deleted{observations} may not detect HI at the flat part of the rotation curve for all galaxies, especially those with $v_{\mathrm{circ}} \leq$  60 km s$^{-1}$. Similarly, a small fraction of low-mass galaxies might not have enough gas to have been detected by HIPASS\deleted{, although this should not impact the VF significantly}. We treat the lowest VF end with caution, and exclude HIPASS datapoints below 60 km s$^{-1}$ from the combined fit. Therefore, the joint velocity function should be representative in the velocity range of 60 $< v_{\mathrm{circ}} < $ 320 km s$^{-1}$.

\subsection{Comparison with simulations}
\label{sec:sims}

We compare our work with a number of simulations.
Shown in Fig. \ref{fig:combined_vcirc_func} are the VFs from the Millennium \citep{Springel2005} and Bolshoi \citep{Klypin2011} dark matter simulations. Here we are plotting friends-of-friends DM halos using two different halo circular velocity definitions: virial velocity $v_{\mathrm{vir}}$ and maximum circular velocity $v_{\mathrm{max}}$. In addition, we show the Illustris-1 DM-only run $v_{\mathrm{max}}$-based VF, constructed for haloes with $M_{\mathrm{DM}}> $ 10$^{10} M_{\mathrm{\odot}}$. 

We also include two VFs measured from Illustris-1 full-physics simulations \citep{Vogelsberger2014, Illustris_coev2014}. Illustris $v_{\mathrm{max}}$ is calculated for all subhaloes with stellar masses $M_{\mathrm{*}} >$  10$^{8} M_{\mathrm{\odot}}$, while Illustris $v_{\mathrm{80}}$ is calculated as the gravitational potential-induced circular rotation velocity at the 80\% stellar mass radius. 

It is strikingly evident that the observed VF does not agree with the dark matter-only simulations, even though the low velocity end of Bolshoi and Millenium simulations displays \deleted{a} marginal agreement with the observational data. At intermediate velocities the dark matter-only VFs sit well below both the observed data and the baryonic simulation. 

However, we find that the observed VF cannot be reconciled with the Illustris $v_{\mathrm{80}}$ and $v_{\mathrm{max}}$-based VFs, \deleted{al}though the full physics simulations produce VFs that are significantly closer to the observed VF. The lack of observed galaxies is evident for velocities lower than $v_{\mathrm{circ}}$ $\approx$ 120 km s$^{-1}$. \replaced{The latter }{This }fact was already shown in \citet{Gonzalez2000, Papastergis2011, Abramson2014} and \citetalias{Zwaan2010}, however, we find it worthy to revisit their results using the latest hydrodynamical simulation results. 

Interestingly, at the intermediate velocities the predicted VFs are systematically offset from the observations, differing by up to a factor of 3. This discrepancy is not related to the "under-abundance" problem.

The mismatch between simulations and observations is either a result of an inconsistency in the way that observations and simulations are measuring the velocity function, or the structure of simulated galaxies is inconsistent with the structure of observed galaxies\deleted{ as measured through the velocity function}. We have not yet performed a fully fair comparison \deleted{of $v_{\mathrm{80}}$} between the simulation\added{s} and observations using the 80\% light radius in both cases, employing adequate surface brightness cuts and including projection effects for the simulation. \added{In the very recent paper by \citet{Maccio2016} it was shown that at least some of the tension 
between the data and models at the low end of the velocity function can be alleviated by considering finite extent of HI disks and relatively larger vertical velocity dispersion. Observational confirmation of their result would go a long way towards explaining the tension in the VF comparison at low circular velocities. We additionally note that while the Illustris VF does not fully match the observed data at high circular velocities, further study of the effects of baryons on the masses and structure of dark matter halos may close the remaining gap.} \replaced{However, this is unlikely to explain the tension in
the VF comparison, which should be considered}{The difference between observed and simulated VF should be considered} to be a constraint on the future generations of galaxy formation models.

\section{Conclusions}
In this work we measure the CALIFA stellar VF, derived from a sample of 226 \deleted{IFS} stellar velocity fields. To our best knowledge, it is the first directly measured VF that includes early-type fast rotators as well as late-types. 

We then combined this VF with the HIPASS VF to obtain the first directly measured velocity function that simultaneously covers a wide range of circular velocities and morphological types. This has the benefit of using the space density and velocity data measured from the same surveys, without assuming scaling relations or conversions between kinematic observables. The combined VF is complete in the range of 60 $< v_{\mathrm{circ}} < $ 320 km s$^{-1}$. We find that Illustris simulation VF does not reproduce the observed data in both the low and high velocity ranges.

The differences between $\Lambda$CDM predictions and the observed VF are not so dissimilar to those found when comparing the halo and stellar mass functions. There, physical processes important for galaxy formation cause a decoupling of the halo and galaxy growth. By highlighting similar discrepancies, this work opens a new window for comparison with theory that should deepen our understanding of galaxy evolution. The resulting velocity function is expected to provide constraints on galaxy assembly and evolution models, insights into \deleted{the} baryonic angular momentum, help improve halo occupation distribution and semi-analytic disk formation models.

\acknowledgments
The authors would like to express gratitude to Louis Abramson for kindly providing their VF data and for assistance with interpreting it. We sincerely thank the referee for the insightful comments.

SB acknowledges support from BMBF through the Erasmus-F project (grant number 05 A12BA1) and is grateful to Leibniz-Institut f\"ur Astrophysik Potsdam for its hospitality during her guest stay there in 2016. JFB acknowledges support from grant AYA2013-48226-C3-1-P from the Spanish Ministry of Economy and Competitiveness (MINECO). CJW acknowledges support through the Marie Curie Career Integration Grant 303912. SFS thanks the CONACYT-125180 and DGAPA-IA100815 projects for
providing him support in this study. DO thanks the University of Western Australia for its
support via a Research Collaboration Award. KS acknowledges funding from the Natural Sciences and Engineering Research Council of Canada.

This study makes use of the data provided by the Calar Alto Legacy Integral Field Area (CALIFA) survey
(http://www.califa.caha.es). Based on observations collected at the Centro Astronòmico Hispano Alem\'{a}n (CAHA) at Calar Alto, operated jointly by the Max-Planck-Institut für Astronomie and the Instituto de Astrofisica de Andalucia (CSIC). CALIFA is the first legacy survey being performed at Calar Alto.

We have extensively used open source data analysis and visualisation tools \textit{Matplotlib} \citep{Matplotlib} and \textit{SciPy} \citep{SciPy}. 
\bibliographystyle{aasjournal}

\listofchanges
\end{document}